\documentclass[12pt]{article}
\usepackage{amssymb,amsmath,amsfonts,eurosym,geometry,ulem,graphicx,caption,color,setspace,sectsty,comment,footmisc,caption,natbib,pdflscape,subfigure,array}
\usepackage[colorlinks=true,linkcolor=blue,urlcolor=blue,citecolor=blue,anchorcolor=blue]{hyperref}
\usepackage{authblk}

\usepackage[table]{xcolor} 
\usepackage{booktabs} 
\usepackage[framemethod=TikZ]{mdframed}
\newcolumntype{L}[1]{>{\raggedright\let\newline\\arraybackslash\hspace{0pt}}m{#1}}
\newcolumntype{C}[1]{>{\centering\let\newline\\arraybackslash\hspace{0pt}}m{#1}}
\newcolumntype{R}[1]{>{\raggedleft\let\newline\\arraybackslash\hspace{0pt}}m{#1}}
\date{}
\begin{document}
\begin{titlepage}
\title{Stress index strategy enhanced with financial news sentiment analysis for the equity markets}
\author[1, 3]{Baptiste Lefort}
\author[2, 3]{Eric Benhamou}
\author[3]{Jean-Jacques Ohana}
\author[3]{David Saltiel}
\author[3]{Beatrice Guez}
\author[3]{Thomas Jacquot}
\affil[1]{CentraleSupélec, Paris-Saclay University}
\affil[2]{Paris-Dauphine PSL}
\affil[3]{Ai For Alpha}
\maketitle
\begin{abstract}
\noindent 
This paper introduces a new \textit{risk-on risk-off} strategy for the stock market, which combines a financial stress indicator with a sentiment analysis done by ChatGPT reading and interpreting Bloomberg daily market summaries. Forecasts of market stress derived from volatility and credit spreads are enhanced when combined with the financial news sentiment derived from GPT4. As a result, the strategy shows improved performance, evidenced by higher Sharpe ratio and reduced maximum drawdowns. The improved performance is consistent across the NASDAQ, the S\&P 500 and the six major equity markets, indicating that the method generalises across equities markets.
\vspace{0.2in}
\vspace{-0.2in}
\bigskip
\end{abstract}
\setcounter{page}{0}
\thispagestyle{empty}
\end{titlepage}
\pagebreak \newpage
\doublespacing

\section{Introduction} \label{sec:introduction}
Recent advancements in Natural Language Processing (NLP) with Large Language Models (LLMs) have made the sentiment analysis of financial news by machines a practical achievement and no longer just a dream. More precisely, Large Language Models (LLMs) have marked a major step forward in processing large contexts, exhibiting human-level performance on various professional and academic benchmarks, although they still have limitations such as reliability issues and limited context windows \citep{openai2023gpt4}. Their ability to process more context has shown particularly interesting applications in many business areas \citep{george2023review}. Hence exploring the potential to extract either weak or strong signals from financial news to enhance a risk-on risk-off investment strategy becomes highly pertinent.

Indeed, extracting sentiment from financial news is not new \citep{Tetlock2007GivingMarket, schumaker2009textual}, and finance has a longstanding tradition of exploiting textual data \citep{kearney2014textual}. 
More precisely, sentiment analysis is a task where we aim at identifying the underlying sentiment in a text. Many different types of models and methods can be employed for identifying sentiment \citep{baccianella-etal-2010-sentiwordnet, tang2023finentity}. 
Models like BERT and its finance-focused version FinBERT have exposed their application in the financial industry \citep{bert_article, finbert_article}. These models have significantly increased the sentiment analysis's precision \citep{kant2018practical}, giving a new opening to using the news for financial decision-making.
Likewise, SentiWordNet 3.0, provides an enhanced lexical resource for sentiment analysis, showing about 20\% accuracy improvement over its earlier version \citep{baccianella-etal-2010-sentiwordnet}. Recent advancements like FinEntity focus on entity-level sentiment classification in financial texts, demonstrating its utility in investment and regulatory applications \citep{tang2023finentity}. Deep Learning application have also shown clear improvement and proved their ability to provide consistently reliable sentiment to a complex text \citep{zhang2018deep}.

However, when scrutinized across various equity markets and extended out-of-sample backtesting, these efforts proved to be less than compelling \cite{xing2018natural}. Interpreting financial news has long been a complex task \citep{loughran2011liability}, as it involves intricate concepts open to various interpretations and contains time-sensitive information with an expiration date of relevance. Moreover, the sentiment conveyed in financial news is often influenced by human perception, and there are numerous underlying implications \citep{ikizlerli2019response,elliott2018negative}. 

We hypothesize two things that can help solve the seminal problem of interpreting news in order to forecast equity market regimes. First, the emergence of Large Language Models (LLMs) could bring fresh perspectives to this longstanding issue, potentially enhancing the interpretation of ambiguities in financial news. Second, news sentiments should be combined with other signals in order to have a robust risk-on risk-off strategy accross major equity markets.

Hence, we present here a new approach to tackle \textit{risk-on, risk-off} strategy for the stock market. This approach integrates a financial stress indicator with sentiment analysis performed by ChatGPT, which reads and interprets daily market summaries from Bloomberg. Additionally, we present a strategy selection method, alternating between the hybrid strategy that combines news signals and stress index and another based only on the conventional stress index indicator.

The rest of the paper is organized as follows. Section \ref{sec:Related Works} provides a review of existing literature on the application of ChatGPT in formulating financial strategies. In section \ref{sec:Key contributions}, we explain how our methodology differs from existing studies and outline our principal contributions. Section \ref{sec:data} presents the different types of data used: namely news, the stress index and investment universe overwhich this new strategy is tested. 
Section \ref{sec:experiments} details the comprehensive set of experiments conducted, highlighting the execution of 18 distinct tests. These tests are designed to ascertain the most effective strategy from a range of alternatives, namely whether the news signal and the stress index signal are effective or not. In total, six varied strategies are evaluated across three disparate financial markets to assert whether the results remains similar across various equity markets.  In Section \ref{sec:results}, we provide an in-depth discussion of the result, with a particular focus on the dynamically strategy that alternates between the stress index and the news sentiment and the stress index alone. The findings reveal that this combined method consistently outperforms other strategies in all three equity markets evaluated. The superiority of this approach is highlighted by its higher Sharpe and Calmar ratio (the ratio of return to maximum drawdown), when compared with the other strategies that rely on the stress index, news alone, the volatility index (VIX) based strategy, and a static combination of stress with news.
Section \ref{sec:conclusion} concludes and suggests future works.

\section{Related Works}\label{sec:Related Works}
In the field of finance and economics, numerous recent academic studies have used ChatGPT, including \citep{Hansen2023CanFedspeak}, \citep{Cowen2023HowGPT}, \citep{Korinek2023LanguageResearch, Lopez-Lira2023CanModels}, and \citep{Noy2023ExperimentalIntelligence}. The capabilities of ChatGPT in various economic and professional contexts are explored in several studies. \citep{Cowen2023HowGPT} and \citep{Korinek2023LanguageResearch} discuss the model's role in economics education and research. \citep{Hansen2023CanFedspeak} focuses on how ChatGPT interprets Fedspeak, the complex language used by the Federal Reserve. In a similar vein, \citep{Lopez-Lira2023CanModels} delves into how to effectively prompt the model for stock return forecasts. Additionally, \citep{Noy2023ExperimentalIntelligence} highlights the model's enhancement of productivity in professional writing, while \citep{Yang2023LargeCredibility} examines its ability to identify credible news sources.
Beside the increasing number of use of LLMs for sentiment analysis, using news sentiment in investment strategy is not new. Many studies involve using more classical methods like lexical approaches for developing news sentiment indicator \citep{9667151}.

The development and utilization of sentiment indicators for trading strategies have also seen significant progress. The study \textit{A Deep Learning Approach with Extensive Sentiment Analysis for Quantitative Investment} by \citep{Wang2023DeepLearning} introduces an approach that incorporates news content for sentiment analysis in quantitative investment, achieving interesting annualized returns. In the realm of cryptocurrency, \citep{Yeoh2023SentimentCryptocurrency} analyze the impact of news sentiment on the prices of cryptocurrencies using the FinBERT model for financial sentiment analysis. Their findings indicate an influence of sentiment on price movements. \citep{Nakagawa2022SentimentLeadLag} also contribute to this field with their investment strategy that exploits the lead-lag effect in company relationships using sentiment analysis. Furthermore, \citep{Yang2023TextMining} demonstrates the profitability of investment strategies based on analyst opinions extracted from text analysis, suggesting an improved daily returns. The market stress is also a widely studied index for combining it with news sentiment indicator and shows promising results \citep{Smales2015RiskOnRiskOff, Smales2016TimeVarying, Lin2023MachineLearningSentiment}. These studies collectively highlight the growing importance and effectiveness of sentiment analysis as a tool for developing investment strategies.

\section{Key contributions}\label{sec:Key contributions}
In this paper, we present a strategy that combines a stress index with news sentiment analysis in a way that has never been addressed. We built a crisis-resistant strategy that yields a constant performance over a long period. Our approach involves several key steps. Firstly, we generate a signal by analyzing news sentiment, utilizing Bloomberg's extensive daily market summaries. This signal provides a broad, market-wide perspective, focusing on impactful financial news. Our findings reveal that using this signal alone is not effective. However, integrating it with the stress index significantly enhances our strategy, making it more consistent and robust. This approach not only improves performance but also demonstrates the broader applicability of the stress index strategy on a macroeconomic level.

The principal contribution of the paper are the following:
\begin{enumerate}
 \item \textbf{News alone is not enough:} Although we use premium and heavily scrutinized data, that is the Bloomberg's daily market reports to create a news sentiment-based macro indicator thanks to a multiple prompt approach using ChatGPT that has demonstrated a statistical correlation with equity market returns, we experienced poor results when using the news sentiment indicator alone. However, when combined with a stress index combining other data type like volatility and credit spread, we get improved performance over our benchmark of a naive long only strategy as well as other strategies based purely on stress index or news.

 \item \textbf{It is crucial to have an alternative method when the news-based signal becomes inefficient:} We present a method to switch between the stress index and the combination of the stress index and the news sentiments based on the persistence of the overall performances of one strategy over the other one. This helps us to mitigate periods of time where the news sentiment indicator becomes inefficient.
 
 \item \textbf{The method works on various equity markets:} Through empirical validation, we confirm that the integration of the stress index and news signal strategy, coupled with the transition to a pure risk index in the event of prolonged underperformance, consistently improves outcomes in various equities markets. This observation underscores the strategy's ability to demonstrate persistent promising results, indicating its capacity to generalize across major equity markets.
\end{enumerate}

\section{Data} \label{sec:data}
To create an investment strategy on various equity markets leveraging financial news and a stress index indicator, we need to collect multiple data: reliable financial news, stress index and investment universe. Let us introduce each of these one at a time.

\subsection{News Data}
\label{sec:bloomberg_daily_market_wraps}
We use professional news from Bloomberg. Bloomberg provides daily market wraps, that are daily summaries of the most important news. They are available since 2010 with growing frequency after 2011. These texts contain both numerical and literal values and are gathering the most informative content and news of the day. Daily market wraps are produced multiple times during the day and specialized for the three regions: the US, Europe and Asia. In order to have a consistent data set, we rely on the latest available news before the European market opening, that is the end of day Asian market wraps. Hence, we collect 3627 daily market wraps. Each of these wraps is about 7000 characters, representing with line breaks about 140 lines or about 5 pages long. Here is below an example of Bloomberg Daily Market Wraps for the day. The range of the contents are any news that may impact financial markets including equity, currencies, cryptocurrencies, bonds and commodities. \\

\vspace{0.2cm}
\begin{mdframed}[linecolor=white, innerleftmargin=10pt, innerrightmargin=10pt]
\textit{- Investor sentiment remains quite negative in China despite a rally in global stocks during the past two months of 2023, Nomura Group analysts including Chetan Sethin Singapore wrote in a client note. In China, there have been more signs of support for the economy, but equity investors still do not appear convinced.}\\
\textit{- Bond Traders Seize on 4\% Yields, Confident Fed Rate Cuts Coming}\\
\textit{- The Australian dollar fell 0.2\% to \$0.6701}
\textit{...}
\end{mdframed}
\vspace{0.2cm}

\subsubsection{News signal}
We follow the same methodology as in \citep{lefort2024chatgpt}, and use a two-step approach to break the sentiment analysis into simpler subtasks for which chatGPT is better suited, namely text summary and keywords identification. Here are the steps we follow:
\begin{enumerate}
 \item First, we collect the daily market summaries from Bloomberg to stay updated on financial trends. 
 \item Next, we ask ChatGPT to generate 15 notable headlines every day, ensuring we capture the most significant events.
 \item Once we have our headlines, we take a moment to assess their tone, deciding whether they convey positive, negative, or indecisive sentiment. Hence, we calculate the daily count of news evaluated in each category, resulting in a numerical score for each day. It is a simple scale: -1 points for a negative sentiment, 0 for indecisive, and +1 for the the days that are positively charged. 
 \item With these scores in hand, we don't just look at one day in isolation; we add them up across a 10-day period to get a broader view of the market's mood. By averaging these daily scores over 10 days, we're able to smooth out the ups and downs and pinpoint the general trend in news sentiment. 
 \item We then apply a statistical method called z-scoring to this 10-day sentiment average, which helps us understand how strongly the news is leaning compared to the norm.
 \item The final transformation involves calculating the mean of the z-scored signal over the preceding 10 days.
 \item Finally, we obtain a simple binary news indicator. If the z-score shows a positive trend, we set it up as 1, indicating a thumbs-up for market sentiment. If the trend dips negative, we set it to 0, signaling a more cautious outlook. This binary indicator serves as a quick reference to gauge the overall sentiment in the financial news, allowing us to make more informed decisions based on the prevailing mood in the market.
\end{enumerate}

\subsection{Stress Index} \label{sec:stress_index_indicator}
We rely on the stress index as presented in \citep{guilleminot2014financial} to capture contagion effects in the market and anticipates future crisis. The indicator is documented to be predictive since it aims at detecting period of high volatility. We prefer this financial stress indicator to the VIX as it has numerous advantages in detecting market contagion:
\begin{itemize}
 \item \textbf{Comprehensive Risk Analysis}: The stress index combines a broad range of market prices of risk, including CDS contracts, providing a more detailed view of market stress compared to the VIX, which focuses mainly on S\&P 500 index options.
 \item \textbf{Enhanced Normalization Techniques}: It employs z-score based normalization, facilitating more effective comparability across various markets, thereby allowing for a nuanced understanding of stress levels in different market segments.
 \item \textbf{Detection of Market Contagion}: It is specially designed to capture interconnectedness and contagion in financial markets, reflecting the complex dynamics often missed by other indicators like the VIX.
 \item \textbf{Detection of Crises}: Incorporating CDS contracts from the main banks, insurances and governments enables it to capture signals related to crises.
\end{itemize}

\subsubsection{Stress Signal computation}
The step to compute the stress index are the following:
\begin{enumerate}
 \item We start by gathering market data for a variety of assets, which includes indicators like the VIX index, TED spread, and CDS index, along with the volatility data for major equity, bond, and commodity markets.
 
 \item Next, we calculate the price of risk for each asset by standardizing the data with z-scoring, which adjusts for the variability and scales the risk price in terms of its deviation from the mean.
 
 \item Then, we aggregate these z-scores by their respective categories such as equities, emerging bonds, government bonds, financial stocks, foreign exchange, commodities, interest rates, and corporate credit to form category-specific stress indicators.
 
 \item We take the average of these category-specific stress indicators to compile a comprehensive stress index that reflects the overall market conditions.

 \item Finally, we scale the resulting average to fall between 0 and 1 by applying the cumulative distribution function (norm.cdf) to the computed stress index, which normalizes our final index value.
\end{enumerate}

Because the stress index final result is a number between 0 and 1 thanks to the cumulative distribution function of the normal distribution, we directly get a stress index signal. 

\subsection{Markets Tested} \label{subsec:Markets Tested}
We test the strategy on six different equity markets: NASDAQ, S\&P500, Nikkei, Euro Stoxx, and Emerging Markets from January 2005 to 2024. 

We compare the strategy on three cases: 
\begin{enumerate}
    \item The S\&P500 alone.
    \item The NASDAQ alone.
    \item An equally weighted basket of the aforementioned six equity markets. 
\end{enumerate}

\section{Experiments} \label{sec:experiments}
\subsection{Transaction Costs}\label{sec:transactioncosts}
To have very realistic simulations, we incorporate linear transaction costs, denoted by $b$ equals to 2 basis points for all our strategy ($b = 0.0002$). To put some mathematical notations, assuming we have $n$ assets whose weights are denoted by $(w_t^i)_{i=1,...,n}$ and that the daily returns of our assets are denoted by $(r_t^i)_{i=1,...,n}$, we compute our strategy value recursively with the convention that the strategy starts at 1 at time 0: $S_{0}=1$ as follows:
\begin{equation}
S_{t} = S_{t-1} \times (1+ \sum_{i=1}^n w_{t-1}^i \times r_t^i - b | w_{t} - w_{t-1}|)
\end{equation}

\subsection{The six different strategies}
In order to evaluate whether the news signal is predictive or not, we compare six strategies and run in total eighteen experiments as we compare these strategies over 3 different markets, namely the US market by testing these strategies on the NASDAQ, the S\&P 500, and an equally weighted strategy combining the 6 major equity markets presented in section \ref{subsec:Markets Tested}.

Here is the list of the 6 different strategies:
\begin{enumerate}
 \item Long only: also referred to as the Benchmark in our comparison figures, this strategy takes a long only position on the tested markets. We term this the Benchmark because our aim is to assess whether the other quantitative strategies exhibit superior performance compared to this one.

 \item VIX: Weights are directly related to the VIX signal where we assume that periods of stress are identified by a VIX level above its 80 percentile which is around 26 percents.
 
 \item SI for Stress Index: Weights are directly related to the stress index signal.
 
 \item News: Weights are directly related to the news signal
 
 \item SI News: Weights are obtained by the straighforward multiplication of the two signals, namely the stress index times the news signal.
 
 \item Dynamic SI News: this strategy is a dynamic selection between the strategy based on stress index with news and the strategy using the stress index alone. More details can be found in the section \ref{subsec:Dynamic Strategy Selection}
\end{enumerate}

In total we have 18 experiments as we compare these 6 strategies over the three cases listed below:
\begin{enumerate}
 \item Test on the NASDAQ.
 \item Test on the S\&P 500.
 \item Test on the 6 major world equity markets.
\end{enumerate}

Also to compare these strategies with a naive long only strategy, we also compute a long only strategy that shares the same volatility as the better performing strategy. Hence, we calculate ex-post the volatility of our best strategies and rescale the benchmark to have the same volatility to compare track records with the benchmark in a volatility consistent way.

\subsection{Dynamic Strategy Selection}\label{subsec:Dynamic Strategy Selection}
The primary goal involves dynamically shifting between two investment strategies: one only reliant on the stress index (SI) and another that incorporates news signals alongside the stress index (SI+News). This strategic alternation aims to navigate efficiently through periods where the inclusion of news signals doesn't substantially improve the performance of the strategy compared to using the stress index alone. Empirical observations reveal a consistent pattern: the SI+News strategy either significantly outperforms or underperforms the SI-only strategy. Consequently, a strategic selector mechanism has been developed. This mechanism computes the Sharpe ratio (defined as the ratio of the excess return over the annualized volatility as in \citep{Sharpe_1966,Sharpe_1975}) over a span of 250 observations, equivalent to an annual rolling period, for the preceding month and consequently selects the strategy that demonstrated superior performance for the forthcoming month. Notably, there are intervals where the SI-only strategy surpasses the combined SI+News strategy. The news signals predominantly act as a performance enhancer during specific scenarios, particularly in times of crisis.

To elucidate this strategic selection, refer to Table \ref{tab:selection}. For each month, the Sharpe ratios of the two strategies are calculated. The strategy for the subsequent month is chosen based on the higher Sharpe ratio. For example, in December 2022, the SI+News strategy, with a Sharpe ratio of 0.9, surpasses the SI-only strategy, which has a Sharpe ratio of 0.4. Consequently, for December, January, and February, the dominant SI+News strategy is selected. Conversely, in March and April, the SI-only strategy becomes dominant, leading to its selection in April and May.

\begin{table}[!htbp]
  \centering
  \caption{Illustration of the Method for Switching Between the SI and SI+News Strategies}
  \begin{tabular}{llll}
  \toprule
  \textbf{Month} & \textbf{Sharpe SI} & \textbf{Sharpe SI+News} & \textbf{Selected Strategy} \\
  \midrule
  Dec 2022 & 0.4 & 0.9 & \ldots \\
  Jan 2023 & -0.1 & 0.7 & SI+News \\
  Feb 2023 & 0.2 & 0.5 & SI+News \\
  Mar 2023 & 0.5 & 0.1 & SI+News  \\
  Apr 2023 & 1.2 & 0.6 & SI \\
  May 2023 & \ldots & \ldots & SI \\
  \ldots & \ldots & \ldots & \ldots \\
  \bottomrule
  \end{tabular}%
  \label{tab:selection}%
\end{table}%

To validate the predictive efficacy of each strategy, it is necessary to demonstrate that the strategy selection exhibits a degree of persistence. The primary indicator of predictive capability is the frequency of selection for each strategy. Over the period from 2011 to 2024, the SI-only strategy was selected 71\% of the time, while the SI+News strategy was chosen 29\% of the time. A random selection mechanism would yield a 50\% selection rate for each strategy. Additionally, a graphical representation of the strategy selection demonstrates that the SI+News strategy is chosen primarily during specific periods, suggesting that the news signal is particularly beneficial during certain times.

\section{The results}\label{sec:results}
We test our six different strategies from January 2005 to January 2024, hence testing them over 19 years. Our comparative analysis, as detailed in tables \ref{tab:SP500}, \ref{tab:Nasdaq}, and \ref{tab:World}, highlights some consistency in the comparison of the different investment strategies for the S\&P 500, NASDAQ, and major equity markets with the uniform outperformance of the Dynamic SI+News strategy in terms of Sharpe and Calmar ratios across all markets. It underscores its superior risk-adjusted returns and drawdown management. In the three tables, in order to have short column name, we denote Sharpe for the Sharpe ratio, Calmar  for the Calmar ratio, Vol for the annualized volatility and Max DD for the maximum drawdowns. We recall that the Calmar ratio is defined as the annualized return over the maximum drawdowns (see \citep{Young_1991}).

\begin{table}[!htbp]
  \centering
  \caption{Comparative Analysis of Investment Strategies for the S\&P 500}
  \begin{tabular}{lrrrrrr}
  \toprule
    {\textbf{Strategy}} &   {\textbf{ Sharpe }} &   {\textbf{ Calmar }} &   {\textbf{Vol}} &   {\textbf{Max DD}} & \textbf{ Turnover } \\
  \midrule
  Dynamic SI+News &  0.81  &  0.56  & 7.5\%  & 11\%  &  8.6  \\
  SI  &  0.70  &  0.51  & 8.0\%  & 11\%  &  7.7  \\
  SI+News &  0.53  &  0.30  & 6.2\%  & 11\%  &  13.4  \\
  Long Only &  0.45  &  0.13  & 7.5\%  & 27\%  & {n.a.} \\
  VIX  &  0.42  &  0.17  & 8.9\%  & 22\%  &  18.5  \\
  News  &  0.42  &  0.15  & 10.6\% & 29\%  &  17.9  \\
  \end{tabular}%
  \label{tab:SP500}%
\end{table}%

In more details, in the S\&P 500 (Table \ref{tab:SP500}), the Dynamic SI+News strategy achieves not only the highest Sharpe ratio (0.81) but also the highest Calmar ratio (0.56). This improvement of Sharpe ratio is quite important as the benchmark strategy (the long only strategy) only achieves a Sharpe ratio of 0.45, making the Sharpe ratio improvement quite substantial. Similarly, in the NASDAQ market (Table \ref{tab:Nasdaq}), the same strategy shows a remarkable Sharpe ratio of 0.89, further reinforcing its efficacy in a tech-heavy index. The analysis for the major equity markets (Table \ref{tab:World}) also corroborates the superiority of the Dynamic SI+News strategy, particularly in managing market downturns, as evidenced by its lower maximum drawdown (Max DD) compared to other strategies like News and VIX. 

If we refine our analysis on the Sharpe ratio, we can also notice that the strategy based on the Stress index alone always comes second indicating that the signals emitted by the stress index seem quite robust and more effective than the ones using the VIX index. Regarding turnover, which measures the frequency of trading within the portfolio, we observe notable differences across strategies. For instance, in the S\&P 500 (Table \ref{tab:SP500}), the 'SI+News' strategy exhibits the highest turnover rate at 13.4, indicating a more active trading approach. This contrasts with strategies like 'SI' and 'Dynamic SI News' which have lower turnover rates, suggesting a more passive strategy. The buy and hold strategy has by definition no turnover as we hold the position for ever. It is interesting to note that the Stress Index based strategy is effectively more moderate in terms of turning the position compared to the VIX based strategy. 

\begin{table}[!htbp]
  \centering
  \caption{Comparative Analysis of Investment Strategies for the NASDAQ}
  \begin{tabular}{lrrrrrr}
  \toprule
    {\textbf{Strategy}} &   {\textbf{ Sharpe }} &   {\textbf{ Calmar }} &   {\textbf{Vol}} &   {\textbf{Max DD}} & \textbf{ Turnover } \\
  \midrule
  Dynamic SI+News &  0.89  &  0.62  & 9.2\%  & 13\%  &  8.6  \\
  SI  &  0.84  &  0.53  & 9.8\%  & 16\%  &  7.6  \\
  Long Only &  0.62  &  0.20  & 9.2\%  & 28\%  & {n.a.} \\
  SI+News &  0.61  &  0.38  & 7.5\%  & 12\%  &  13.4  \\
  VIX  &  0.58  &  0.25  & 11.3\% & 27\%  &  18.5  \\
  News  &  0.43  &  0.15  & 12.3\% & 34\%  &  17.9  \\
  \end{tabular}%
  \label{tab:Nasdaq}%
\end{table}%

\begin{table}[!htbp]
  \centering
  \caption{Comparative Analysis of Investment Strategies for the major equity markets}
  \begin{tabular}{lrrrrrr}
  \toprule
    {\textbf{Strategy}} &   {\textbf{ Sharpe }} &   {\textbf{ Calmar }} &   {\textbf{Vol}} &   {\textbf{Max DD}} & \textbf{ Turnover } \\
  \midrule
  Dynamic SI+News &  0.85  &  0.44  & 6.8\%  & 13\%  &  8.5  \\
  SI  &  0.71  &  0.39  & 7.2\%  & 13\%  &  7.7  \\
  SI+News &  0.61  &  0.20  & 5.5\%  & 17\%  &  13.4  \\
  Long Only &  0.52  &  0.13  & 6.8\%  & 28\%  & {n.a.} \\
  VIX  &  0.42  &  0.15  & 8.4\%  & 23\%  &  18.4  \\
  News  &  0.32  &  0.09  & 8.1\%  & 29\%  &  16.2  \\
  \end{tabular}%
  \label{tab:World}%
\end{table}%

Similarly, in the NASDAQ (Table \ref{tab:Nasdaq}) and major equity markets (Table \ref{tab:World}), the 'VIX' and 'News' strategies are characterized by higher turnover rates (18.5 and 17.9 for NASDAQ, 18.4 and 16.2 for major equity markets, respectively), again reflecting a more active management style. This frequent trading could be indicative of an attempt to capitalize on short-term market movements.

In terms of the Calmar ratio, which assesses the risk-adjusted performance of an investment by comparing the average annual compounded rate of return and the maximum drawdown, the 'Dynamic SI+News' strategy consistently outperforms others across all markets. This is evident from its higher Calmar ratios (0.56 for S\&P 500, 0.62 for NASDAQ, and 0.44 for major equity markets). The superior Calmar ratios suggest that this strategy not only provides higher returns but also does so with less risk, as indicated by lower maximum drawdowns.

In contrast, strategies like 'Long Only' and 'News' show lower Calmar ratios, implying that they bear a higher risk (as seen in their higher maximum drawdowns). This could be a critical factor for investors who are risk-averse and prefer strategies that limit potential losses.

\begin{figure}[!htbp]
 \centering
 \includegraphics[width=\linewidth]{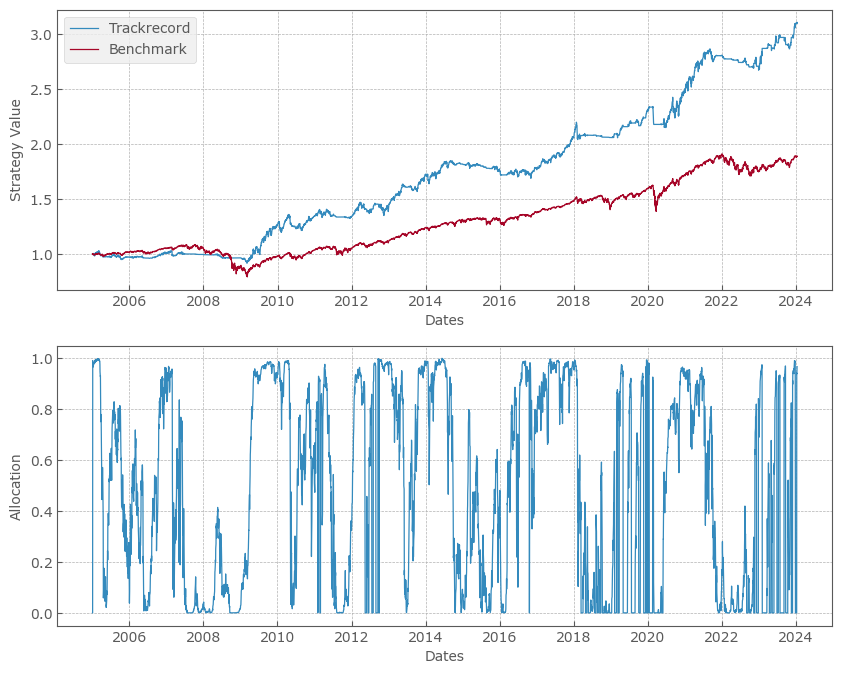}
 \caption{Comparison of the dynamic strategy combining Stress index and News and Stress index alone using persistence of overperformance of one strategy over the other one versus the naive long only strategy (rescaled at the same volatility) for the S\&P 500 universe. The first plot compares the two strategies over time while the subplot provides the corresponding overall allocation.}\label{fig:Dynamic_SI_News_USLarge}%
\end{figure}

\begin{figure}[!htbp]
 \centering
 \includegraphics[width=\linewidth]{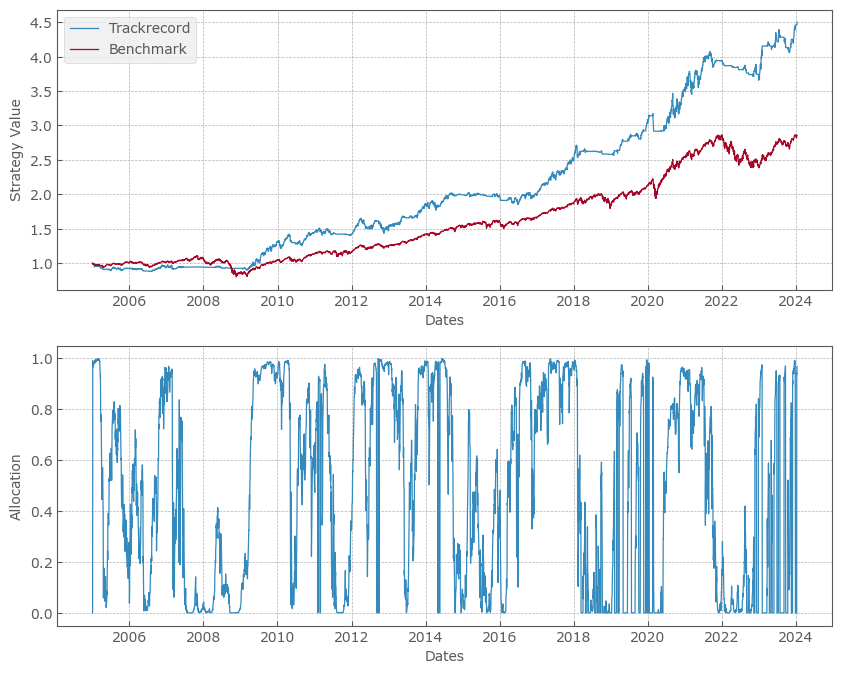}
 \caption{Comparison of the dynamic strategy combining Stress index and News and Stress index alone using persistence of overperformance of one strategy over the other one versus the naive long only strategy (rescaled at the same volatility) for the NASDAQ 100 universe. The first plot compares the two strategies over time while the subplot provides the corresponding overall allocation.}\label{fig:Dynamic_SI_News_USTech}%
\end{figure}

\begin{figure}[!htbp]
 \centering
 \includegraphics[width=\linewidth]{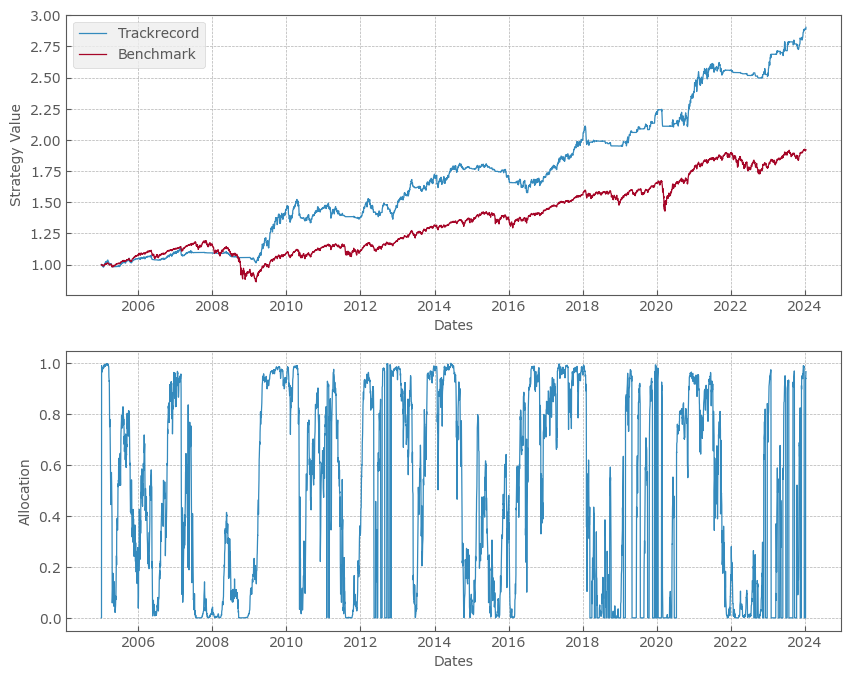}
 \caption{Comparison of the dynamic strategy combining Stress index and News and Stress index alone using persistence of overperformance of one strategy over the other one versus the naive long only strategy (rescaled at the same volatility) for the Major Equities Markets. The first plot compares the two strategies over time while the subplot provides the corresponding overall allocation.}\label{fig:Dynamic_SI_News_WORLD}%
\end{figure}

Figures \ref{fig:Dynamic_SI_News_USLarge}, \ref{fig:Dynamic_SI_News_USTech}, and \ref{fig:Dynamic_SI_News_WORLD} provide a visual representation of these findings and compare the best performing strategy namely the 'Dynamic SI+News' strategy versus the passive long only benchmark. The subplot in all figures \ref{fig:Dynamic_SI_News_USLarge}, \ref{fig:Dynamic_SI_News_USTech}, and \ref{fig:Dynamic_SI_News_WORLD} reveal how the Dynamic SI+News strategy dynamically adjusts its allocation, contributing to its robust performance. 

All in all, when compiling our 18 experiments, our analysis demonstrates the effectiveness of the Dynamic SI News strategy across different markets, with its ability to adapt to market conditions and maintain superior performance metrics standing out as a key takeaway.

\begin{figure}[!htbp]
 \centering
 \includegraphics[width=\linewidth]{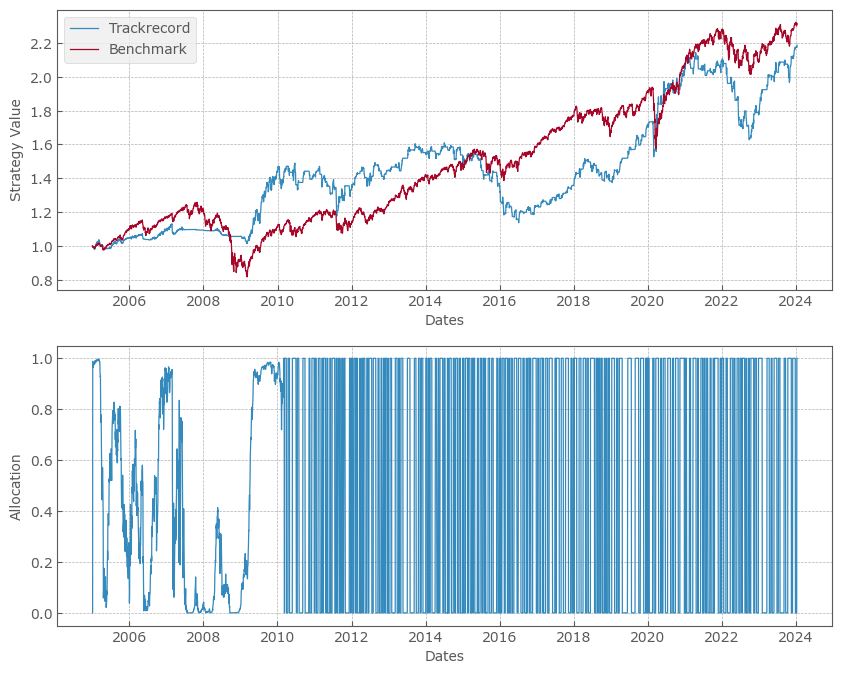}
 \caption{Comparison of the strategy based only on news versus the naive long only strategy (rescaled at the same volatility) for the Major Equities Markets. The first plot compares the two strategies over time while the subplot provides the corresponding overall allocation.}\label{fig:News_WORLD}%
\end{figure}

\begin{figure}[!htbp]
 \centering
 \includegraphics[width=\linewidth]{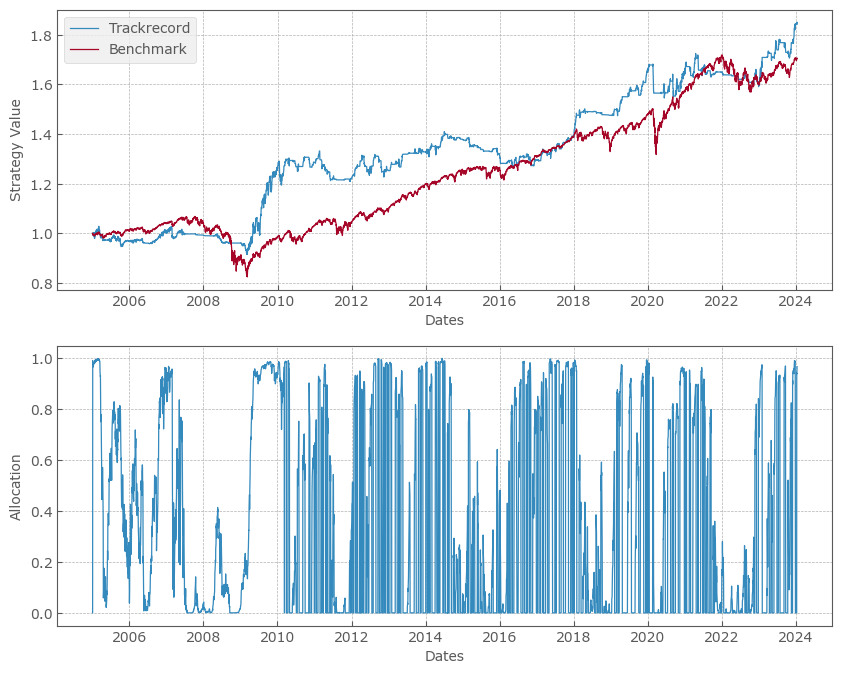}
 \caption{Comparison of the strategy combining Stress index and News versus the naive long only strategy (rescaled at the same volatility) for the S\&P 500 universe. The first plot compares the two strategies over time while the subplot provides the corresponding overall allocation.}\label{fig:NewsAndSI_USLarge}%
\end{figure}

\begin{figure}[!htbp]
 \centering
 \includegraphics[width=\linewidth]{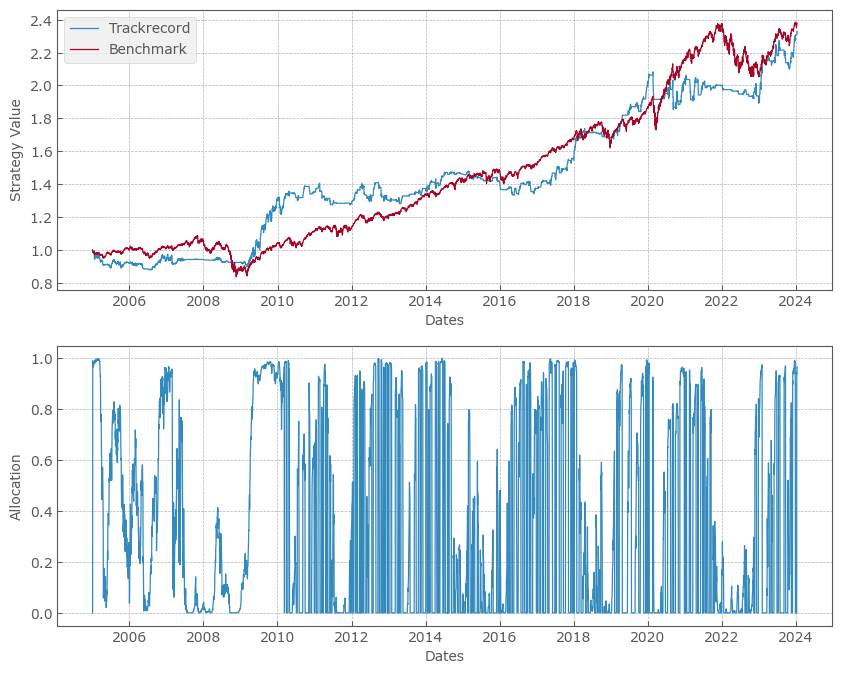}
 \caption{Comparison of the strategy combining Stress index and News versus the naive long only strategy (rescaled at the same volatility) for the NASDAQ 100 universe. The first plot compares the two strategies over time while the subplot provides the corresponding overall allocation.}\label{fig:NewsAndSI_USTech}%
\end{figure}

\begin{figure}[!htbp]
 \centering
 \includegraphics[width=\linewidth]{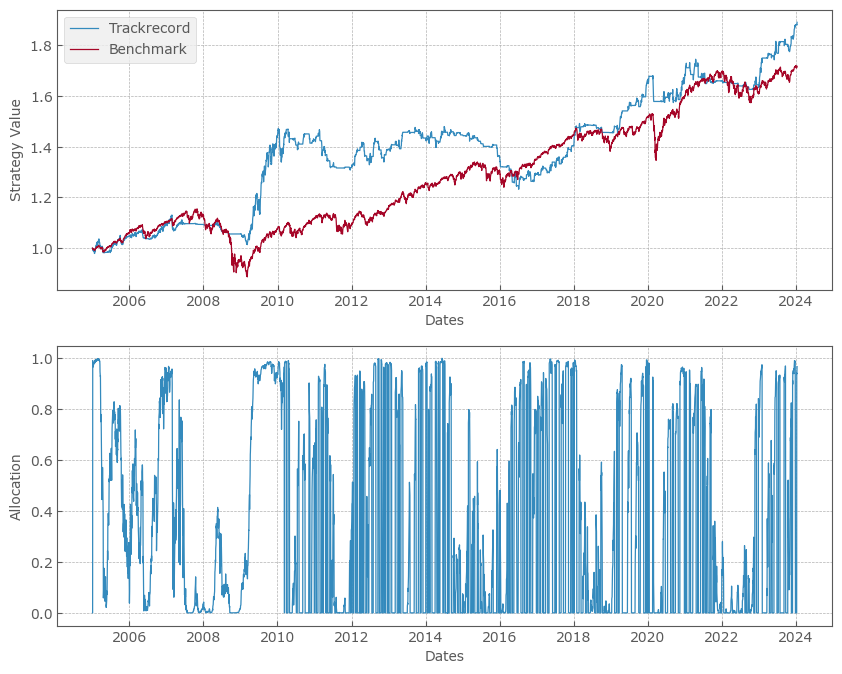}
 \caption{Comparison of the strategy combining Stress index and News versus the naive long only strategy (rescaled at the same volatility) for the Major Equities Markets. The first plot compares the two strategies over time while the subplot provides the corresponding overall allocation.}\label{fig:NewsAndSI_WORLD}%
\end{figure}

\begin{figure}[!htbp]
 \centering
 \includegraphics[width=\linewidth]{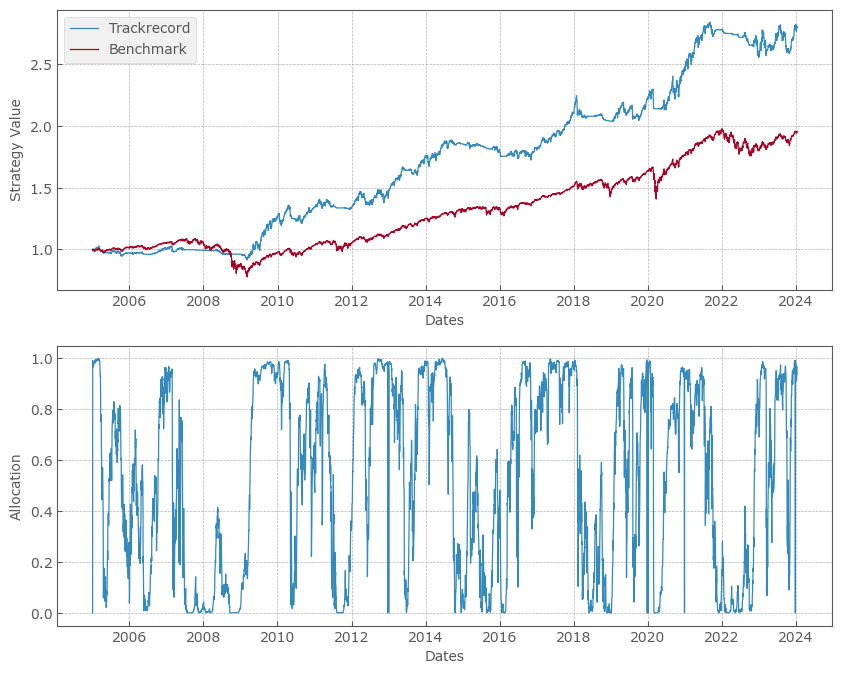}
 \caption{Comparison of the Stress Index based strategy versus the naive long only strategy (rescaled at the same volatility) for the S\&P 500 universe. The first plot compares the two strategies over time while the subplot provides the corresponding overall allocation.}\label{fig:SI_USLarge}%
\end{figure}

\begin{figure}[!htbp]
 \centering
 \includegraphics[width=\linewidth]{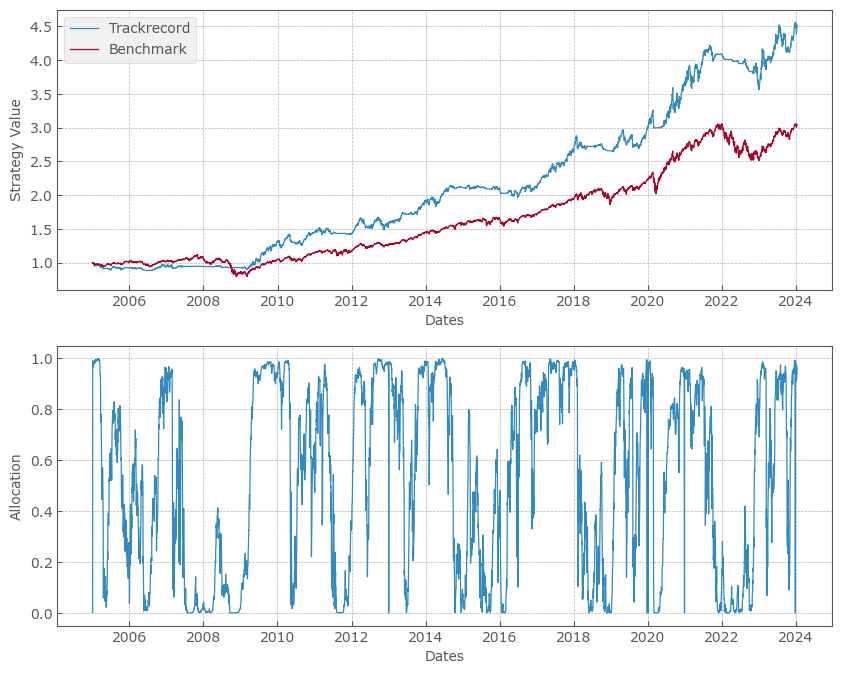}
 \caption{Comparison of the Stress Index based strategy versus the naive long only strategy (rescaled at the same volatility) for the NASDAQ 100 universe. The first plot compares the two strategies over time while the subplot provides the corresponding overall allocation.}\label{fig:SI_USTech}%
\end{figure}

\begin{figure}[!htbp]
 \centering
 \includegraphics[width=\linewidth]{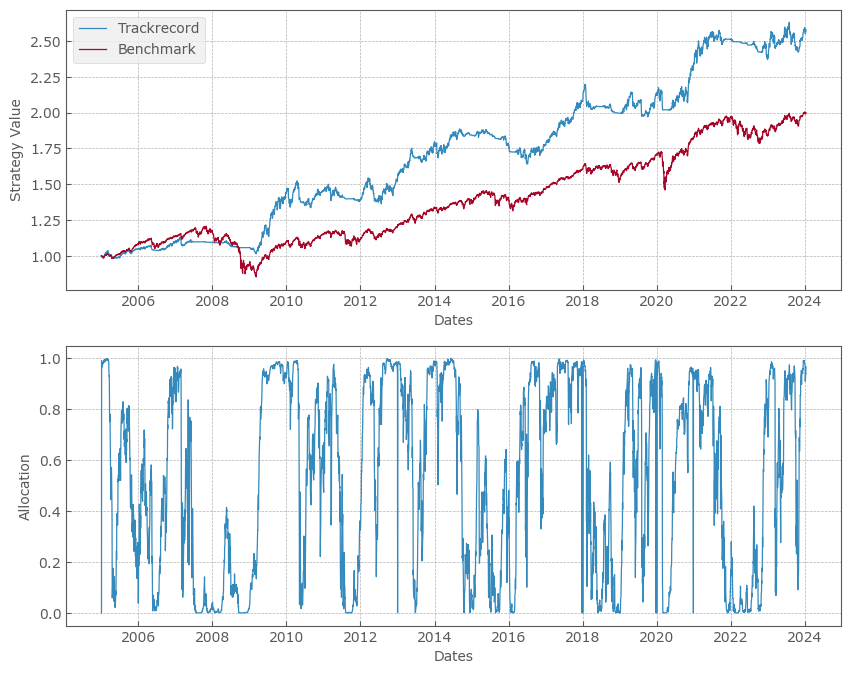}
 \caption{Comparison of the Stress Index based strategy versus the naive long only strategy (rescaled at the same volatility) for the Major Equities Markets. The first plot compares the two strategies over time while the subplot provides the corresponding overall allocation.}\label{fig:SI_WORLD}%
\end{figure}

\begin{figure}[!htbp]
 \centering
 \includegraphics[width=\linewidth]{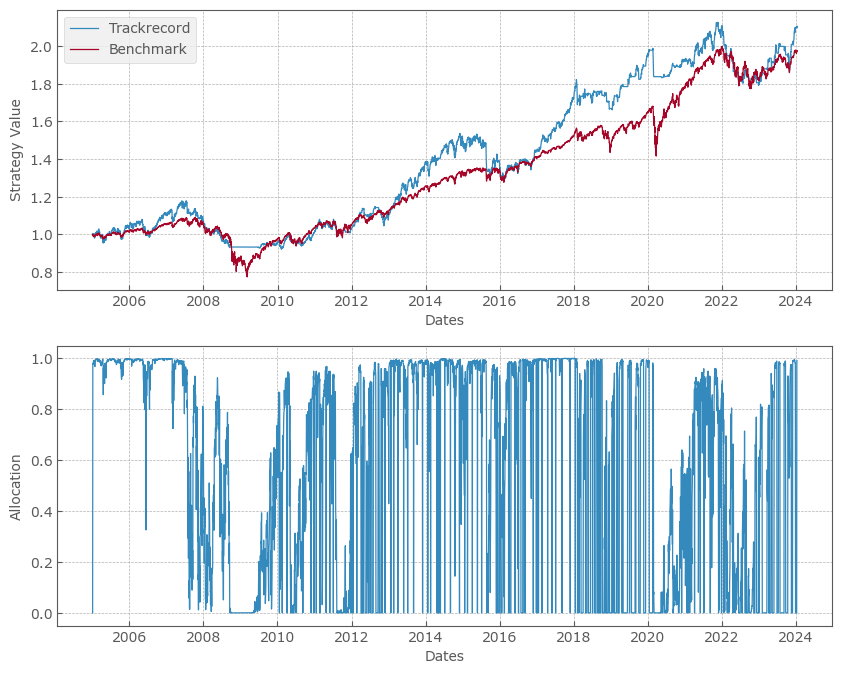}
 \caption{Comparison of the VIX based strategy versus the naive long only strategy (rescaled at the same volatility) for the S\&P 500 universe. The first plot compares the two strategies over time while the subplot provides the corresponding overall allocation.}\label{fig:VIX_USLarge}%
\end{figure}

\begin{figure}[!htbp]
 \centering
 \includegraphics[width=\linewidth]{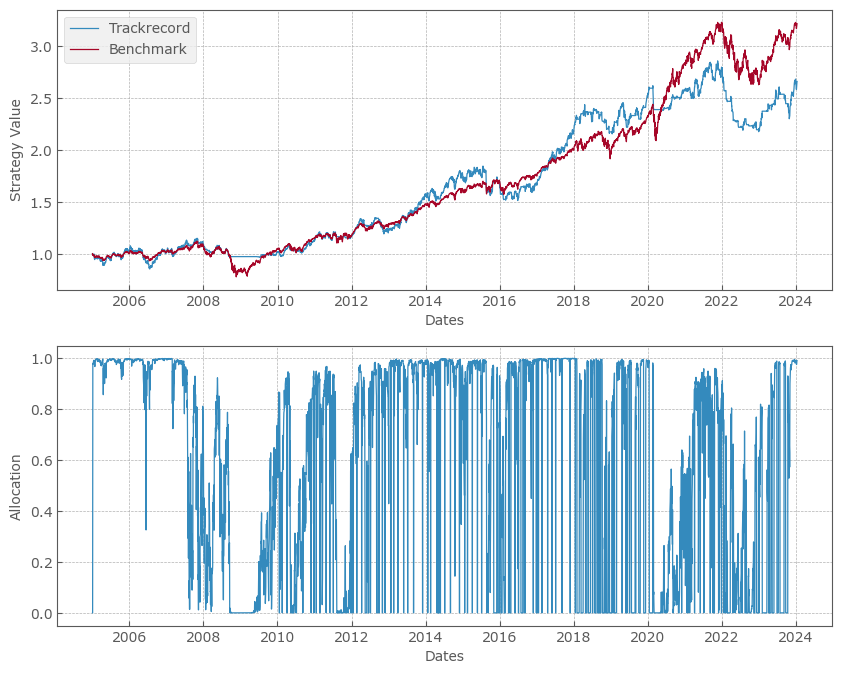}
 \caption{Comparison of the VIX based strategy versus the naive long only strategy (rescaled at the same volatility) for the NASDAQ 100 universe. The first plot compares the two strategies over time while the subplot provides the corresponding overall allocation.}\label{fig:VIX_USTech}%
\end{figure}

\begin{figure}[!htbp]
 \centering
 \includegraphics[width=\linewidth]{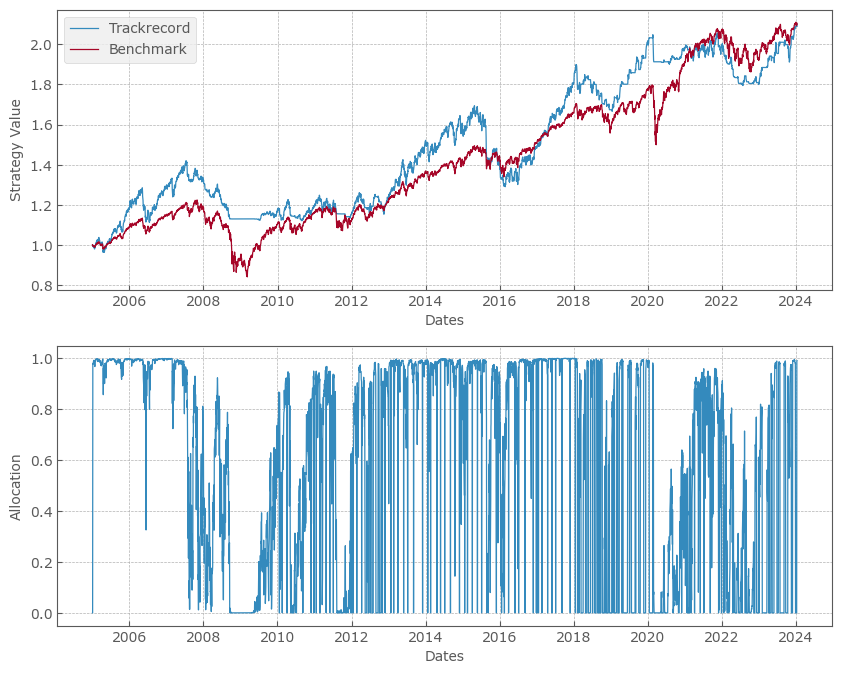}
 \caption{Comparison of the VIX based strategy versus the naive long only strategy (rescaled at the same volatility) for the Major Equities Markets. The first plot compares the two strategies over time while the subplot provides the corresponding overall allocation.}\label{fig:VIX_World}%
\end{figure}

Additionally we provide in the appendix figures \ref{fig:News_USLarge} \ref{fig:News_USTech}, \ref{fig:News_WORLD}, \ref{fig:NewsAndSI_USLarge}, \ref{fig:NewsAndSI_USTech}, \ref{fig:NewsAndSI_WORLD}, \ref{fig:SI_USLarge}, \ref{fig:SI_USTech}, \ref{fig:SI_WORLD}, \ref{fig:VIX_USLarge}, \ref{fig:VIX_USTech}, \ref{fig:VIX_World} the  comparative performance of other investment strategies over time, alongside their corresponding allocations and over the 3 different market universes, namely the S\&P 500, the NASDAQ, and Major Equities Markets. For each figure, we give as the top plot the temporal performance comparison with the benchmark, the long only strategy, while the subplot details the overall allocation.

Figure \ref{fig:News_USLarge} illustrates the performance of a news-based strategy against a naive long-only strategy within the S\&P 500 universe. This figure highlights the temporal evolution of the strategy's effectiveness in comparison to the benchmark. In Figure \ref{fig:News_USTech}, we observe a similar analysis within the NASDAQ universe. The subplot is particularly insightful for understanding the allocation differences underlined by the news-based strategy. Figure \ref{fig:News_WORLD} extends this comparison to the Major Equities Markets, offering a broader view of the strategy's applicability in a global context.

Figures \ref{fig:NewsAndSI_USLarge}, \ref{fig:NewsAndSI_USTech}, and \ref{fig:NewsAndSI_WORLD} delve into strategies that amalgamate news with stress index data. These figures offer an intriguing perspective on the synergistic effects of combining these two data sources across different market universes.

Figures \ref{fig:SI_USLarge}, \ref{fig:SI_USTech}, and \ref{fig:SI_WORLD} focus on strategies purely based on the stress index. These plots are crucial for understanding the standalone impact of the stress index on the investment strategy.

Finally, Figures \ref{fig:VIX_USLarge}, \ref{fig:VIX_USTech}, and \ref{fig:VIX_World} compare the VIX-based strategy with the naive long-only approach. The temporal comparison and allocation subplots underscore the unique characteristics and performance of the VIX-based strategy in different market settings.

\subsection{Intuition of the results}
The integration of news signals with the stress index in investment strategies, as demonstrated in the S\&P 500, NASDAQ, and major equity markets (Tables \ref{tab:SP500}, \ref{tab:Nasdaq}, and \ref{tab:World}), makes a lot of sense and is grounded in both empirical results and theoretical underpinnings from financial literature.

\subsubsection{Added Values of News}
Intuitively incorporating news read by a machine into an investment strategy based on the stress index can improve the strategy's effectiveness for several reasons:
\begin{itemize}
\item \textbf{Signal Reactivity:} News can detect new trend or sentiment in the markets quicker. This is effectively illustrated by the higher turnover of the pure news strategy.

\item \textbf{Objectivity and Consistency:} Unlike human analysts, reading news by a machine is less prone to cognitive biases or emotional responses. The news signal should be quite consistent and objective.

\item \textbf{Quality of the data:} Since we use Bloomberg market wraps that are supposed to capture all major news, the news signal should be triggered by any weak signals from news.

\item \textbf{Hidden pattern catching:} The news signal is capable of capturing more complex relationships between news items than a human can.
\end{itemize}

\subsubsection{Empirical Evidence}
As already presented, the Dynamic SI+News strategy, which combines both news signals and stress index data, shows a remarkable performance across various markets. In the S\&P 500, this strategy not only achieves the highest Sharpe ratio (0.81) but also a significant Calmar ratio (0.56), surpassing the benchmark long only strategy's Sharpe ratio of 0.45.

\subsubsection{Theoretical Insights}
Combining news signals with the stress index aligns with the principles of Behavioral Finance, particularly as detailed in \citep{barber2008all}. Their research demonstrates how news and media attention significantly influence investor behavior, often leading to irrational decision-making based on recent news rather than long-term trends. This indicates a potential edge in combining news analysis with stress indices, which act as barometers for market sentiment and stress. By integrating recent news, which potentially triggers overreactions or underreactions, with the more steady and sentiment-focused stress indices, a more balanced and informed investment the strategy can be improved by capitalizing on the quick, news-driven shifts in investor behavior while being anchored by the broader, more stable market insights provided by the stress index.

\subsubsection{Role of Stress Index}
The stress index alone consistently ranks second in terms of Sharpe ratio across different markets, underscoring its robustness. It acts as a gauge of market tension and uncertainty, often signaling impending market corrections or volatility, making it a crucial component of a holistic investment strategy.

\subsubsection{Turnover Analysis}
Analyzing turnover rates sheds light on strategy aggressiveness. The 'SI+News' strategy exhibits the highest turnover (13.4 in the S\&P 500), indicative of an active trading approach. This is quite logical as the news based signals is supposed to react very quickly. The same applies to the VIX strategy that seems to overreact to any high volatility environement. In contrast, 'SI' and 'Dynamic SI+News' strategies demonstrate more moderation, suggesting a balanced approach that leverages the predictive power of stress indices while avoiding excessive trading.

\subsubsection{Impact on Maximum Drawdown and Calmar Ratio}
The integration of news signals into the stress index strategy has a significant impact on the maximum drawdown (Max DD) and, consequently, on the Calmar ratio. This is evident in the comparative analysis of investment strategies across various markets (Tables \ref{tab:SP500}, \ref{tab:Nasdaq}, and \ref{tab:World}).

\paragraph{Max DD Reduction}
The Dynamic SI+News strategy consistently demonstrates a lower Max DD compared to other strategies, especially the News-only strategy. For example, in the S\&P 500, the Max DD for Dynamic SI+News is 11\%, markedly lower than the 29\% Max DD for the News strategy. This reduction in Max DD is crucial for risk-averse investors and indicates enhanced strategy stability during market downturns.

\paragraph{Enhanced Calmar Ratio}
The Calmar ratio, which measures the return per unit of downside risk, is notably higher in the Dynamic SI+News strategy across all markets. For instance, in the S\&P 500, this strategy achieves a Calmar ratio of 0.56, surpassing the Stress Index (SI) alone at 0.51 and significantly outperforming the News strategy at 0.15. This trend is consistent across the NASDAQ and major equity markets.

\paragraph{Rationale for Improvement}
Incorporating news into the stress index strategy enhances its ability to adapt to market sentiments and trends rapidly. News provides real-time insights and immediate market reactions, which, when combined with the stress index's broader market sentiment gauge, allows for a more dynamic and responsive strategy. This combination effectively mitigates risks during volatile periods, leading to a reduced Max DD and an improved Calmar ratio.

\paragraph{Literature Support}
This finding aligns with the existing financial literature that emphasizes the importance of combining various types of market information to achieve a more comprehensive and robust investment strategy. For instance, \citep{Tetlock2007GivingMarket} highlights how media and news content significantly influence investor sentiment and stock market dynamics, underscoring the value of incorporating real-time news into investment strategies \citep{Tetlock2007GivingMarket}. Similarly, \cite{baker2016measuring} in their study provide insights into how economic policy uncertainty, derived from news media coverage, impacts market conditions and investor behavior, further justifying the inclusion of news data alongside stress index information in strategic decision-making\citep{Tetlock2007GivingMarket}. Additionally, \cite{da2011search} through their work emphasize the influence of media attention on stock prices and trading volumes, suggesting that the attention garnered by specific news can be pivotal in financial market movements and should be integrated into comprehensive investment strategies \citep{da2011search}. These studies collectively support the synergy between real-time news and stress index data, enhancing the ability to capture and react to market anomalies and stress conditions, thereby improving the overall risk-adjusted performance of the strategy.

\section{Conclusion}\label{sec:conclusion}
This paper introduces a novel \textit{risk-on, risk-off} strategy for the stock market, leveraging a financial stress indicator combined with sentiment analysis performed by ChatGPT on Bloomberg's daily market summaries. The strategy enhances market stress forecasts, which are based on volatility and credit spreads, using financial news sentiment derived from GPT4. This results in a significant improvement in performance, evidenced by a higher Sharpe ratio and reduced maximum drawdowns. The method's effectiveness is not limited to a single market; it is consistent across various equity markets, including the NASDAQ and S\&P 500, as well as six other major equity markets, indicating its broad applicability. There are many potential directions for future research. Future works could investigate if the strategy can be applied to different financial markets such as commodities, bonds, and foreign exchange markets, to test its versatility and robustness. 
Additionally, it would be worthwhile to investigate whether additional data sources, such as social media sentiment or macroeconomic indicators, could further improve the strategy's performance. Identifying patterns in economic events where the news signal is most effective is also a potential avenue for future research.

\section{Bibliographical References}
\bibliographystyle{plainnat}
\bibliography{main}

\begin{thebibliography}{37}
\providecommand{\natexlab}[1]{#1}
\providecommand{\url}[1]{\texttt{#1}}
\expandafter\ifx\csname urlstyle\endcsname\relax
  \providecommand{\doi}[1]{doi: #1}\else
  \providecommand{\doi}{doi: \begingroup \urlstyle{rm}\Url}\fi

\bibitem[Araci(2019)]{finbert_article}
Dogu Araci.
\newblock Finbert: Financial sentiment analysis with pre-trained language models.
\newblock \emph{CoRR}, abs/1908.10063, 2019.
\newblock URL \url{http://arxiv.org/abs/1908.10063}.

\bibitem[Baccianella et~al.(2010)Baccianella, Esuli, and Sebastiani]{baccianella-etal-2010-sentiwordnet}
Stefano Baccianella, Andrea Esuli, and Fabrizio Sebastiani.
\newblock {S}enti{W}ord{N}et 3.0: An enhanced lexical resource for sentiment analysis and opinion mining.
\newblock In Nicoletta Calzolari, Khalid Choukri, Bente Maegaard, Joseph Mariani, Jan Odijk, Stelios Piperidis, Mike Rosner, and Daniel Tapias, editors, \emph{Proceedings of the Seventh International Conference on Language Resources and Evaluation ({LREC}'10)}, Valletta, Malta, May 2010. European Language Resources Association (ELRA).
\newblock URL \url{http://www.lrec-conf.org/proceedings/lrec2010/pdf/769_Paper.pdf}.

\bibitem[Baker et~al.(2016)Baker, Bloom, and Davis]{baker2016measuring}
Scott~R Baker, Nicholas Bloom, and Steven~J Davis.
\newblock Measuring economic policy uncertainty.
\newblock \emph{The quarterly journal of economics}, 131\penalty0 (4):\penalty0 1593--1636, 2016.

\bibitem[Barber and Odean(2008)]{barber2008all}
Brad~M Barber and Terrance Odean.
\newblock All that glitters: The effect of attention and news on the buying behavior of individual and institutional investors.
\newblock \emph{The review of financial studies}, 21\penalty0 (2):\penalty0 785--818, 2008.

\bibitem[Cowen and Tabarrok(2023)]{Cowen2023HowGPT}
Tyler Cowen and Alexander~T. Tabarrok.
\newblock {How to Learn and Teach Economics with Large Language Models, Including GPT}.
\newblock \emph{SSRN Electronic Journal}, XXX\penalty0 (XXX):\penalty0 0--0, 3 2023.
\newblock ISSN 1556-5068.
\newblock \doi{10.2139/SSRN.4391863}.
\newblock URL \url{https://papers.ssrn.com/abstract=4391863}.

\bibitem[Da et~al.(2011)Da, Engelberg, and Gao]{da2011search}
Zhi Da, Joseph Engelberg, and Pengjie Gao.
\newblock In search of attention.
\newblock \emph{The journal of finance}, 66\penalty0 (5):\penalty0 1461--1499, 2011.

\bibitem[Devlin et~al.(2018)Devlin, Chang, Lee, and Toutanova]{bert_article}
Jacob Devlin, Ming{-}Wei Chang, Kenton Lee, and Kristina Toutanova.
\newblock {BERT:} pre-training of deep bidirectional transformers for language understanding.
\newblock \emph{CoRR}, abs/1810.04805, 2018.
\newblock URL \url{http://arxiv.org/abs/1810.04805}.

\bibitem[Elliott et~al.(2018)Elliott, Grant, and Hodge]{elliott2018negative}
W~Brooke Elliott, Stephanie~M Grant, and Frank~D Hodge.
\newblock Negative news and investor trust: The role of \$ firm and \# ceo twitter use.
\newblock \emph{Journal of Accounting Research}, 56\penalty0 (5):\penalty0 1483--1519, 2018.

\bibitem[George and George(2023)]{george2023review}
A~Shaji George and AS~Hovan George.
\newblock {A review of ChatGPT AI's impact on several business sectors}.
\newblock \emph{Partners Universal International Innovation Journal}, 1\penalty0 (1):\penalty0 9--23, 2023.

\bibitem[Guilleminot et~al.(2014)Guilleminot, Ohana, and Ohana]{guilleminot2014financial}
Beno{\^\i}t Guilleminot, Jean-Jacques Ohana, and Steve Ohana.
\newblock A new financial stress indicator: properties and conditional asset price behavior.
\newblock \emph{SSRN Electronic Journal}, January 2014.
\newblock Available at SSRN: \url{https://ssrn.com/abstract=2317321} or \url{http://dx.doi.org/10.2139/ssrn.2317321}.

\bibitem[Hansen and Kazinnik(2023)]{Hansen2023CanFedspeak}
Anne~Lundgaard Hansen and Sophia Kazinnik.
\newblock {Can ChatGPT Decipher Fedspeak?}
\newblock \emph{SSRN Electronic Journal}, XX\penalty0 (XX):\penalty0 XX, 3 2023.
\newblock ISSN 1556-5068.
\newblock \doi{10.2139/SSRN.4399406}.
\newblock URL \url{https://papers.ssrn.com/abstract=4399406}.

\bibitem[Ikizlerli et~al.(2019)Ikizlerli, Holmes, and Anderson]{ikizlerli2019response}
Deniz Ikizlerli, Phil Holmes, and Keith Anderson.
\newblock The response of different investor types to macroeconomic news.
\newblock \emph{Journal of Multinational Financial Management}, 50:\penalty0 13--28, 2019.

\bibitem[Januário et~al.(2022)Januário, Carosia, Silva, and Coelho]{9667151}
Brenda~A. Januário, Arthur Emanuel de~O. Carosia, Ana Estela A.~da Silva, and Guilherme~P. Coelho.
\newblock Sentiment analysis applied to news from the brazilian stock market.
\newblock \emph{IEEE Latin America Transactions}, 20\penalty0 (3):\penalty0 512--518, 2022.
\newblock \doi{10.1109/TLA.2022.9667151}.

\bibitem[Kant et~al.(2018)Kant, Puri, Yakovenko, and Catanzaro]{kant2018practical}
Neel Kant, Raul Puri, Nikolai Yakovenko, and Bryan Catanzaro.
\newblock Practical text classification with large pre-trained language models, 2018.

\bibitem[Kearney and Liu(2014)]{kearney2014textual}
Colm Kearney and Sha Liu.
\newblock Textual sentiment in finance: A survey of methods and models.
\newblock \emph{International Review of Financial Analysis}, 33:\penalty0 171--185, 2014.

\bibitem[Korinek(2023)]{Korinek2023LanguageResearch}
Anton Korinek.
\newblock {Language Models and Cognitive Automation for Economic Research}.
\newblock \emph{Cambridge, MA}, XX\penalty0 (XX):\penalty0 XX, 2 2023.
\newblock \doi{10.3386/W30957}.
\newblock URL \url{https://www.nber.org/papers/w30957}.

\bibitem[Lefort et~al.(2024)Lefort, Benhamou, Ohana, Saltiel, Guez, and Challet]{lefort2024chatgpt}
Baptiste Lefort, Eric Benhamou, Jean-Jacques Ohana, David Saltiel, Beatrice Guez, and Damien Challet.
\newblock {Can ChatGPT Compute Trustworthy Sentiment Scores from Bloomberg Market Wraps}.
\newblock \emph{Available at SSRN}, January 2024.
\newblock URL \url{https://papers.ssrn.com/sol3/papers.cfm?abstract_id=4688451}.

\bibitem[Li et~al.(2023)Li, Hu, and Luo]{Wang2023DeepLearning}
Wang Li, Chaozhu Hu, and Youxi Luo.
\newblock A deep learning approach with extensive sentiment analysis for quantitative investment.
\newblock \emph{Electronics}, 12\penalty0 (18):\penalty0 3960, 2023.
\newblock \doi{10.3390/electronics12183960}.
\newblock URL \url{https://dx.doi.org/10.3390/electronics12183960}.

\bibitem[Lin et~al.(2023)Lin, Yu, Chen, and Huang]{Lin2023MachineLearningSentiment}
Tse-Mao Lin, Jing-Long Yu, Jhih-Wei Chen, and Chin-Sheng Huang.
\newblock Application of machine learning with news sentiment in stock trading strategies.
\newblock \emph{International Journal of Financial Research}, 14:\penalty0 1, 05 2023.
\newblock \doi{10.5430/ijfr.v14n3p1}.

\bibitem[Lopez-Lira and Tang(2023)]{Lopez-Lira2023CanModels}
Alejandro Lopez-Lira and Yuehua Tang.
\newblock {Can ChatGPT Forecast Stock Price Movements? Return Predictability and Large Language Models}.
\newblock \emph{SSRN Electronic Journal}, XXX\penalty0 (XX-XX):\penalty0 XX, 4 2023.
\newblock ISSN 1556-5068.
\newblock \doi{10.2139/SSRN.4412788}.
\newblock URL \url{https://papers.ssrn.com/abstract=4412788}.

\bibitem[Loughran and McDonald(2011)]{loughran2011liability}
Tim Loughran and Bill McDonald.
\newblock When is a liability not a liability? textual analysis, dictionaries, and 10-ks.
\newblock \emph{The Journal of finance}, 66\penalty0 (1):\penalty0 35--65, 2011.

\bibitem[Nakagawa et~al.(2022)Nakagawa, Sashida, and Sakaji]{Nakagawa2022SentimentLeadLag}
Kei Nakagawa, Shingo Sashida, and Hiroki Sakaji.
\newblock Investment strategy via lead lag effect using economic causal chain and ssestm model.
\newblock In \emph{IIAI International Conference on Advanced Applied Informatics}, 2022.
\newblock \doi{10.1109/IIAIAAI55812.2022.00065}.
\newblock URL \url{https://dx.doi.org/10.1109/IIAIAAI55812.2022.00065}.

\bibitem[Noy and Zhang(2023)]{Noy2023ExperimentalIntelligence}
Shakked Noy and Whitney Zhang.
\newblock {Experimental Evidence on the Productivity Effects of Generative Artificial Intelligence}.
\newblock \emph{SSRN Electronic Journal}, XX\penalty0 (XX):\penalty0 XX, 3 2023.
\newblock \doi{10.2139/SSRN.4375283}.
\newblock URL \url{https://papers.ssrn.com/abstract=4375283}.

\bibitem[OpenAI(2023)]{openai2023gpt4}
OpenAI.
\newblock Gpt-4 technical report, 2023.

\bibitem[Schumaker and Chen(2009)]{schumaker2009textual}
Robert~P Schumaker and Hsinchun Chen.
\newblock Textual analysis of stock market prediction using breaking financial news: The azfin text system.
\newblock \emph{ACM Transactions on Information Systems (TOIS)}, 27\penalty0 (2):\penalty0 1--19, 2009.

\bibitem[Sharpe(1966)]{Sharpe_1966}
William~F. Sharpe.
\newblock Mutual fund performance.
\newblock \emph{The Journal of Business}, pages 119--138, January 1966.

\bibitem[Sharpe(1975)]{Sharpe_1975}
William~F. Sharpe.
\newblock Adjusting for risk in portfolio performance measurement.
\newblock \emph{Journal of Portfolio Management}, pages 29--34, Winter 1975.

\bibitem[Smales(2015)]{Smales2015RiskOnRiskOff}
L.A. Smales.
\newblock Risk-on / risk-off: Financial market response to investor fear.
\newblock \emph{Journal of International Financial Markets, Institutions and Money}, 2015.
\newblock School of Economics \& Finance, Curtin University, Australia.

\bibitem[Smales(2016)]{Smales2016TimeVarying}
Lee~A. Smales.
\newblock {Time-varying relationship of news sentiment, implied volatility and stock returns}.
\newblock \emph{Applied Economics}, 48\penalty0 (51):\penalty0 4942--4960, November 2016.
\newblock \doi{10.1080/00036846.2016.116}.
\newblock URL \url{https://ideas.repec.org/a/taf/applec/v48y2016i51p4942-4960.html}.

\bibitem[Tang et~al.(2023)Tang, Yang, Huang, Tam, and Tang]{tang2023finentity}
Yixuan Tang, Yi~Yang, Allen~H Huang, Andy Tam, and Justin~Z Tang.
\newblock Finentity: Entity-level sentiment classification for financial texts, 2023.

\bibitem[Tetlock(2007)]{Tetlock2007GivingMarket}
Paul~C. Tetlock.
\newblock {Giving Content to Investor Sentiment: The Role of Media in the Stock Market}.
\newblock \emph{The Journal of Finance}, 62\penalty0 (3):\penalty0 1139--1168, 6 2007.
\newblock ISSN 1540-6261.
\newblock \doi{10.1111/J.1540-6261.2007.01232.X}.
\newblock URL \url{https://onlinelibrary.wiley.com/doi/full/10.1111/j.1540-6261.2007.01232.x}.

\bibitem[Xing et~al.(2018)Xing, Cambria, and Welsch]{xing2018natural}
Frank~Z Xing, Erik Cambria, and Roy~E Welsch.
\newblock Natural language based financial forecasting: a survey.
\newblock \emph{Artificial Intelligence Review}, 50\penalty0 (1):\penalty0 49--73, 2018.

\bibitem[Yang(2023)]{Yang2023TextMining}
Cheol-Won Yang.
\newblock Investment strategy via analyst report text mining.
\newblock \emph{Journal of Data and Qualitative Science}, 2023:\penalty0 1--12, 2023.
\newblock \doi{10.1108/jdqs-09-2022-0022}.
\newblock URL \url{https://dx.doi.org/10.1108/jdqs-09-2022-0022}.

\bibitem[Yang and Menczer(2023)]{Yang2023LargeCredibility}
Kai-Cheng Yang and Filippo Menczer.
\newblock {Large language models can rate news outlet credibility}.
\newblock Technical report, arxiv, 4 2023.
\newblock URL \url{https://arxiv.org/abs/2304.00228v1}.

\bibitem[Yeoh et~al.(2023)Yeoh, Chung, and Wang]{Yeoh2023SentimentCryptocurrency}
Eik~Den Yeoh, Tinfah Chung, and Yuyang Wang.
\newblock Predicting price trends using sentiment analysis: A study of stepn’s socialfi and gamefi cryptocurrencies.
\newblock \emph{Cryptocurrency Markets}, 44, 2023.
\newblock \doi{10.37256/cm.4420232572}.
\newblock URL \url{https://dx.doi.org/10.37256/cm.4420232572}.

\bibitem[Young(1991)]{Young_1991}
T.W. Young.
\newblock Calmar ratio: A smoother tool.
\newblock \emph{Futures}, 20, Nr 1:\penalty0 40--41, 1991.

\bibitem[Zhang et~al.(2018)Zhang, Wang, and Liu]{zhang2018deep}
Lei Zhang, Shuai Wang, and Bing Liu.
\newblock Deep learning for sentiment analysis: A survey.
\newblock \emph{Wiley Interdisciplinary Reviews: Data Mining and Knowledge Discovery}, 8\penalty0 (4):\penalty0 e1253, 2018.

\end{thebibliography}

\clearpage
\section*{Figures} \label{sec:fig}
\addcontentsline{toc}{section}{Figures}

\begin{figure}[!htbp]
 \centering
 \includegraphics[width=\linewidth]{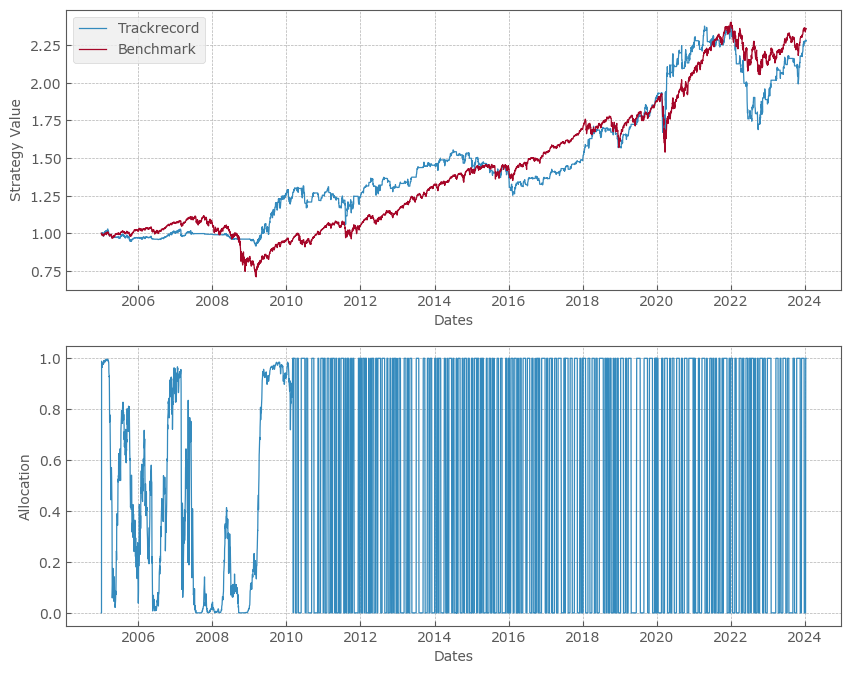}
 \caption{Comparison of the strategy based only on news versus the naive long only strategy (rescaled at the same volatility) for the S\&P 500 universe. The first plot compares the two strategies over time while the subplot provides the corresponding overall allocation.}\label{fig:News_USLarge}%
\end{figure}

\begin{figure}[!htbp]
 \centering
 \includegraphics[width=\linewidth]{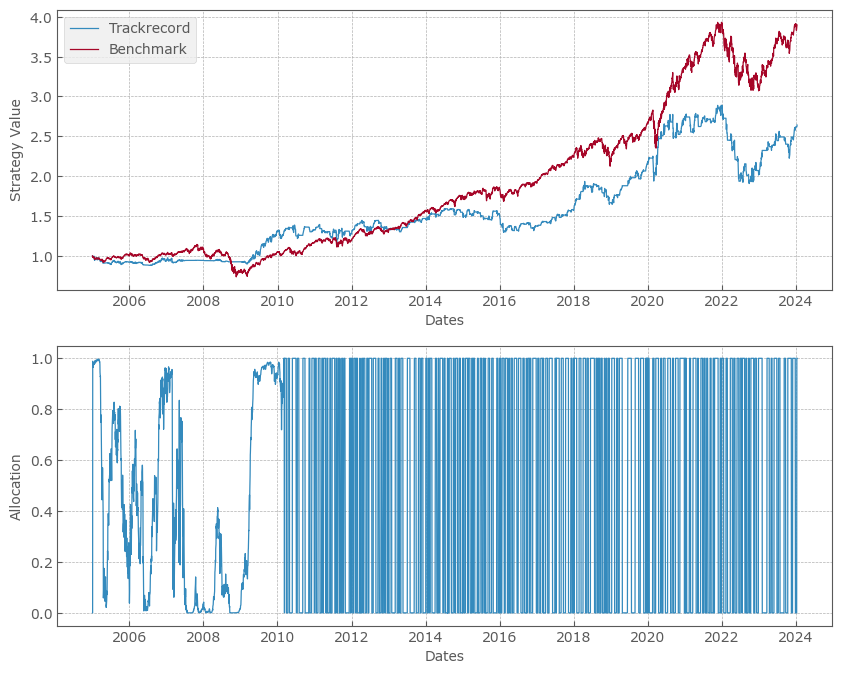}
 \caption{Comparison of the strategy based only on news versus the naive long only strategy (rescaled at the same volatility) for the NASDAQ 100 universe. The first plot compares the two strategies over time while the subplot provides the corresponding overall allocation.}\label{fig:News_USTech}%
\end{figure}

\end{document}